\documentclass{appolb}

\usepackage{tikz}
\usepackage[bf,labelsep=period,skip=0cm]{caption}

\usepackage{multicol}
\usepackage{cancel}
\usepackage{braket}
\usepackage{amsmath}
\usepackage{xcolor}

\usepackage{amssymb}
\usepackage{graphicx}
\usepackage{slashed}
\usepackage{verbatim}
\usepackage{dsfont,latexsym} 
\usepackage{color}

\usepackage{wrapfig}

\newcommand{\be}{\begin{equation}}
\newcommand{\ee}{\end{equation}}
\newcommand{\ba}{\begin{eqnarray}}
\newcommand{\ea}{\end{eqnarray}}
\newcommand{\ban}{\begin{eqnarray*}} 
\newcommand{\ean}{\end{eqnarray*}}
\newcommand \nn {\nonumber}

\def\q{{\mathbf q}}
\def\k{{\mathbf k}}
\def\x{{\mathbf x}}

\def\z{{\mathbf z}}
\def\w{{\mathbf w}}
\def\v{{\mathbf v}}

\def\uv{{\underline{v}}}  

\begin{document}
\title{DIS dijet production at next-to-eikonal accuracy in the CGC%
\thanks{ Presented by Arantxa Tymowska at XXIX Cracow Epiphany Conference on Physics at the Electron-Ion Collider and Future Facilities, Kraków, Poland, 16-19 January 2023.}%
}
\author{Tolga Altinoluk, Guillaume Beuf, Alina Czajka and Arantxa Tymowska
\address{Theoretical Physics Division, National Centre for Nuclear Research, Pasteura 7, Warsaw 02-093, Poland}
\\[3mm]
}
\maketitle
\begin{abstract}
In this work, we derive the cross-section for inclusive DIS dijet production at full next-to-eikonal order. We include the corrections that stem from taking a finite width of the target, the interaction of the quark with the transverse component of the background field and also the dynamics of the target.
\end{abstract}
  
\section{Introduction}
 The results of this work are computed in the framework of the Color Glass Condensate (CGC) effective theory \cite{CGC}. CGC is defined in the limit of very high energies where one keeps the resolution scale fixed (Regge-Gribov limit) and modifies the Bjorken-$x$ in order to make it smaller and smaller, hence it is also called the small-$x$ limit. The CGC formalism is used for the case of high energy dilute-dense scattering where two approximations are taken into account. First, the semi-classical approximation, where the dense target is treated as the classical background field, $A^\mu_a(x)$, and the dilute projectile in this case is taken as a virtual photon treated in perturbation theory. The second approximation is the eikonal approximation, which is discussed in the next section. These proceedings present the results that were obtained in \cite{mipaper2}.

\section{The eikonal approximation and corrections}

As previously mentioned, in the framework of CGC it is very common to adopt the eikonal approximation which is analogous to taking the asymptotically high energy limit $s\to \infty$. In this limit, we boost the target, which develops a hierarchy among the different components of the background field  $A^\mu_a(x)$ and its coordinates:

 \begin{equation}
    A^\mu_a(x)\rightarrow
    \begin{cases}
      \gamma_t A^-_a\big(\gamma_t x^+,\frac{1}{\gamma_t}x^-,\x\big) \\
      \frac{1}{\gamma_t}A^+_a\big(\gamma_t x^+,\frac{1}{\gamma_t}x^-,\x\big)\\
      A^i_a\big(\gamma_t x^+,\frac{1}{\gamma_t}x^-,\x\big)\nn
    \end{cases}
 \end{equation}
 where $\gamma_t$ is the Lorentz boost factor. From this follows that the power-enhanced component is $A^-_a$ and that the hierarchy among coordinates sets $x^+$ as enhanced and the $x^-$ as suppressed. In order to take the eikonal limit, one needs to take three different approximations. The first one is the so-called shockwave approximation and it relies on taking the background field to be localized at $x^+=0$ so we have $A^\mu_a(x) \propto \delta (x^+)$; the second one is to disregard the components of the background field that are not power-enhanced in energy, so  $A^\mu_a(x)\simeq\delta^{\mu -}A^-_a(x)$. Finally, in the last approximation, the power-suppressed coordinate dependence is not taken into account, therefore one finds $A^\mu_a(x)\simeq A^\mu_a(x^+,\x)$. 

If one wants to go to next-to-eikonal (NEik) corrections, one has to relax any of the three aforementioned approximations. Relaxing the first approximation would give a finite width of the target, allowing a transverse motion of the parton within the medium. Relaxing the second approximation, we would need to include the interactions with the perpendicular component of the background field. These corrections have been taken into account in \cite{mipaper, giovanni1,giovanni2}. However, in our most recent work \cite{mipaper2}, we also relaxed the third approximation, therefore including the dependence on the $x^-$ coordinate. The final result is then at full next-to-eikonal order in the gluon background field. These corrections were also implemented in \cite{tolgaguillaume} at the level of the propagator. 

\section{Inclusive DIS dijet production}

The computation of the inclusive DIS dijet production is mostly motivated by the fact that it is one of the processes that the future Electron-Ion Collider (EIC) will focus on. The energies that the EIC will probe are lower when compared to the ones at the Large Hadron Collider (LHC), where the eikonal approximation is used. Therefore, computing the process at next-to-eikonal order would give us corrections in the energy of order $1/s$ that will be of significant importance for the EIC.

In order to compute this process, one needs to take into account the contribution of two types of diagrams. The first diagram corresponds to the splitting of the photon into a quark-antiquark pair before the medium, and it contributes at both eikonal and NEik orders. The second diagram corresponds to the splitting into a quark-antiquark pair inside the medium. This diagram contributes only at NEik order, since, the shockwave approximation has to be relaxed in order in order to compute this diagram. 

\section{Cross-section for inclusive DIS dijet production}



In this section, we present the cross-section for DIS dijet production, where we set $k_1, k_2$ to be the final momenta for the quark and antiquark, and $q$ to be the initial momentum of the photon.
We computed the cross-section for both the longitudinal and the transverse polarization of the photon. The longitudinal polarization amounts to taking  $\epsilon^\lambda_\mu(q) \to \epsilon^L_\mu (q) \equiv \frac{Q}{q^+} g^+_\mu $, where $Q$ is the virtuality of the photon. The transverse polarization is $ \left\{
\epsilon_{\lambda}^+(q)=0, 
\epsilon_{\lambda}^i(q)= \varepsilon_{\lambda}^i ,
\epsilon_{\lambda}^-(q)= \frac{\q^i\varepsilon^i_{\lambda}}{q^+} \right \}.
$ 
 In the case of the transverse polarization of the photon, we get contributions from both diagrams that correspond to the splitting of the photon either before or inside the medium. However, for the longitudinal polarization of the photon, only the diagram corresponding to the splitting of the photon before the medium gives non-vanishing contributions.
In \cite{mipaper2} we computed the complete set of propagators and amplitudes taking into account all the aforementioned corrections. This led to the following decorations on Wilson lines that we can find in the cross-section:

{\footnotesize \begin{align}
\mathcal{U}^{(1)}_{F;j} ( \v) =&\,  \int_{-\frac{L^+}{2}}^{\frac{L^+}{2}}dv^+\, 
\mathcal{U}_F\Big(\frac{L^+}{2},v^+;\v\Big) 
\overleftrightarrow{\mathcal{D}_{\v^j}}
\mathcal{U}_F\Big(v^+,-\frac{L^+}{2};\v\Big) \nn
\\
\mathcal{U}^{(2)}_F ( \v) =&\,
\int_{-\frac{L^+}{2}}^{\frac{L^+}{2}}dv^+\,  
\mathcal{U}_F\Big(\frac{L^+}{2},v^+;\v\Big) 
\overleftarrow{\mathcal{D}_{\v^j}}\, \overrightarrow{\mathcal{D}_{\v^j}} 
\mathcal{U}_F\Big(v^+,-\frac{L^+}{2};\v\Big) \nn
\\
\mathcal{U}^{(3)}_{F; ij} ( \v)
 =&\,  
\int_{-\frac{L^+}{2}}^{\frac{L^+}{2}}dv^+\,  
\mathcal{U}_F\Big(\frac{L^+}{2},v^+;\v\Big) 
gt \!\cdot\! \mathcal{F}_{ij}(\uv) 
\mathcal{U}_F\Big(v^+,-\frac{L^+}{2};\v\Big) \nn
\end{align}}
where $L^+$ is the finite width of the target.

\subsection{NEik DIS dijet production cross-section via longitudinal photon}

The cross-section for the longitudinal polarization of the photon is the sum of the generalized eikonal result and the explicit next-to-eikonal corrections. The so-called generalized eikonal result is when the Wilson lines depend on the $b^-$ coordinate. If this $b^-$ is set to be zero, one would recover the strict eikonal result. This generalized eikonal cross-section expressed in terms of dipoles ($d$) and quadrupoles ($Q$) is given by:
{\footnotesize \begin{align}
\frac{d\sigma_{\gamma^{*}_L\rightarrow q_1\bar q_2}}{d {\rm P.S.}}\Bigg|_{\rm Gen. \, Eik}
&= N_c\, \frac{\alpha_{\textrm{em}} }{\pi}\, e_f^2\, Q^2\, 
\theta\Big(q^+\!+\!k_1^+\!-\!k_2^+\Big) \theta\Big(q^+\!-\!k_1^+\!+\!k_2^+\Big) 
\frac{k_1^+k_2^+}{(q^+)^5}\;
(q^+\!+\!k_1^+\!-\!k_2^+)^2 (q^+\!-\!k_1^+\!+\!k_2^+)^2  
\nn \\
&
\times
\int_{\v, \v', \w, \w'}
e^{i \k_1 \cdot (\v'-\v) } \, e^{i\k_2 \cdot (\w'-\w) } 
K_0(\hat{Q}|\w'-\v'|) K_0(\hat{Q}|\w-\v|)
 \nn \\
&
\times
\int d (\Delta b^-) e^{i\Delta b^-(k_1^++k_2^+ - q^+)}
\bigg\{
Q\Big(\w',\v',\v,\w,\frac{\Delta b^-}{2}\Big)
-d\big(\w',\v'\big)-d\big(\v,\w\big)+1
\bigg\}\nn
\end{align} }
Where $\alpha_{em}=e^2/(4\pi)$, $K_0$ is the modified Bessel function and $\hat{Q} = \sqrt{m^2 +  \frac{(q^+\!+\!k_1^+\!-\!k_2^+)(q^+\!-\!k_1^+\!+\!k_2^+)}{4(q^+)^2}\, Q^2}$. The corrections are presented in terms of quadrupoles  ( $\tilde Q,Q^{(1)}_j$ and $Q^{(2)}$) and dipoles ($\tilde d,d^{(1)}$ and $d^{(2)}$) defined in \cite{mipaper2}. These decorated quadrupoles and dipoles include the different decorations on the Wilson lines. The explicit beyond eikonal corrections in the cross-section are:
{\footnotesize \begin{align}
 \frac{d\sigma_{\gamma^{*}_L\rightarrow q_1\bar q_2}}{d {\rm P.S.}}\Bigg|_{\rm NEik \, corr.}^{\textrm{dec. on }q}
&
=   2\pi \delta(k_1^+\!+\!k_2^+\!-\!q^+) 
\;   8  N_c\,  \frac{\alpha_{\textrm{em}}}{\pi} \,e_f^2\,
 Q^2\,   
 \frac{(k_1^+)^2 (k_2^+)^3}{(q^+)^5}\,
  \nn \\
&
\times
  2 {\rm Re} \int_{\v, \v', \w, \w'}
e^{i \k_1 \cdot (\v'-\v) } \, e^{i\k_2 \cdot (\w'-\w) } 
\textrm{K}_0\left(\bar{Q}\, |\w'\!-\!\v'|\right)
\textrm{K}_0\left(\bar{Q}\, |\w\!-\!\v|\right)
 \nn \\
&
\hspace{-3cm}\times
 \Bigg\{
 \Big[\frac{ (\k_2^j\!-\!\k_1^j)}{2}\,
+\frac{i}{2}\,  \partial_{\w^j} \Big]
 \Big[Q^{(1)}_j(\w',\v',\v_*,\w)-d^{(1)}_j(\v_*,\w)\Big]
 -i \Big[Q^{(2)}(\w',\v',\v_*,\w)-d^{(2)}(\v_*,\w)\Big]
  \Bigg\}
\nn
\end{align}}

{\footnotesize \begin{align}
 \frac{d\sigma_{\gamma^{*}_L\rightarrow q_1\bar q_2}}{d {\rm P.S.}}\Bigg|_{\rm NEik \, corr.}^{\textrm{dec. on }\bar{q}}
&
=   2\pi \delta(k_1^+\!+\!k_2^+\!-\!q^+) 
\;   8  N_c\,  \frac{\alpha_{\textrm{em}}}{\pi} \,e_f^2\,
 Q^2\,   
 \frac{(k_1^+)^3 (k_2^+)^2}{(q^+)^5}\,
 \nn \\
 &
 \times
2 {\rm Re} \int_{\v, \v', \w, \w'}
e^{i \k_1 \cdot (\v'-\v) } \, e^{i\k_2 \cdot (\w'-\w) } 
\textrm{K}_0\left(\bar{Q}\, |\w'\!-\!\v'|\right)
\textrm{K}_0\left(\bar{Q}\, |\w\!-\!\v|\right)
\nn \\
&
\hspace{-3cm}\times
 \Bigg\{
 \Big[-\frac{(\k_2^j\!-\!\k_1^j)}{2} 
 +\frac{i}{2}\,\partial_{\v^j}
 \Big]
 \Big[{Q^{(1)}_j(\v',\w',\w_*,\v)}^{\dag}-{d^{(1)}_j(\w_*,\v)}^{\dag} \Big]
 -i  \Big[{Q^{(2)}(\v',\w',\w_*,\v)}^{\dag}-{d^{(2)}(\w_*,\v)}^{\dag}\Big]
   \Bigg\}
\nn
\end{align}}

{\footnotesize \begin{align}
 \frac{d\sigma_{\gamma^{*}_L\rightarrow q_1\bar q_2}}{d {\rm P.S.}}\Bigg|_{\rm NEik \, corr.}^{\textrm{dyn. target}}
&
=  2\pi \delta(k_1^+\!+\!k_2^+\!-\!q^+) 
\;   8  N_c\,  \frac{\alpha_{\textrm{em}}}{\pi} \,e_f^2\,
 Q^2\,   
 \frac{(k_1^+)^2 (k_2^+)^2(k_1^+\!-\!k_2^+)}{(q^+)^5}\,
  2 {\rm Re} \, ( -i)\!\!
\int_{\v, \v', \w, \w'}\!\!\!\!\!\!
e^{i \k_1 \cdot (\v'-\v) } \,   
\nn \\
 &
 \hspace{-4.5cm}
\times   e^{i\k_2 \cdot (\w'-\w) }\,
\bigg[\tilde Q(\w',\v',\v_*,\w_*)-\tilde d(\v_*,\w_*) \bigg]
\,    
\textrm{K}_0\left(\bar{Q}\, |\w'\!-\!\v'|\right) \;
 \left[
\textrm{K}_0\left(\bar{Q}\, |\w\!-\!\v|\right)
-\frac{\left(\bar{Q}^2\!-\!m^2\right)}{2\bar{Q}}\, |\w\!-\!\v|\, 
\textrm{K}_1\left(\bar{Q}\, |\w\!-\!\v|\right)
\right]
\nn
\end{align}} 

where $\bar{Q} \equiv \sqrt{m^2 + Q^2 \frac{k_1^+k_2^+}{(q^+)^2}}$.

\subsection{NEik DIS dijet production cross-section via transverse photon}

For the case of the transverse polarization of the photon, the cross-section is divided into a generalized eikonal part and explicit NEik corrections. The generalized eikonal cross section is
{\footnotesize \begin{align}
\frac{d\sigma_{\gamma^{*}_T\rightarrow q_1\bar q_2}}{d {\rm P.S.}}\Bigg|_{\rm Gen. \, Eik}
=&\,
N_c\,  \frac{\alpha_{\textrm{em}}}{\pi} \, e_f^2\,
   \frac{2k_1^+ k_2^+}{q^+}\;
 \theta(q^+\!+\!k_1^+\!-\!k_2^+)\,
 \theta(q^+\!+\!k_2^+\!-\!k_1^+)\,
 \int_{\v, \v', \w, \w'}
e^{i \k_1 \cdot (\v'-\v) } \, e^{i\k_2 \cdot (\w'-\w) }
\nn \\
&
\hspace{-3.5cm}\times
\Bigg\{
2\, m^2\, 
\textrm{K}_0\left(\hat{Q}\, |\w'\!-\!\v'|\right)\, \textrm{K}_0\left(\hat{Q}\, |\w\!-\!\v|\right)
+
\left[1+\left(\frac{k_2^+\!-\!k_1^+}{q^+}\right)^2\right]\,
\hat{Q}^2\,  \frac{(\w'\!-\!\v')\!\cdot\!(\w\!-\!\v)}{|\w'\!-\!\v'||\w\!-\!\v|}\, 
\textrm{K}_1\left(\hat{Q}\, |\w'\!-\!\v'|\right)\,
 \nn \\
&
\hspace{-3.5cm}\times\textrm{K}_1\left(\hat{Q}\, |\w\!-\!\v|\right)
\Bigg\}
\int d (\Delta b^-) e^{i\Delta b^-(k_1^++k_2^+ - q^+)} 
\bigg\{
Q\Big(\w',\v',\v,\w,\frac{\Delta b^-}{2}\Big)
-d\big(\w',\v'\big)-d\big(\v,\w\big)+1
\bigg\} \nn
\end{align}}

and the explicit beyond eikonal corrections are:
{\footnotesize \begin{align}
& \frac{d\sigma_{\gamma^{*}_T\rightarrow q_1\bar q_2}}{d {\rm P.S.}}\Bigg|_{\rm NEik \, corr.}^{\textrm{in}}
=\,
  2\pi \delta(k_1^+\!+\!k_2^+\!-\!q^+)\,  
N_c\, \alpha_{\textrm{em}}\,    e_f^2 \,  
  \left[1+\left(\frac{k_2^+\!-\!k_1^+}{q^+}\right)^2\right]\, 
2 {\rm Re}\;  (i)
 \int_{\z, \v', \w'}
e^{i \k_1 \cdot (\v'-\z) } \, e^{i\k_2 \cdot (\w'-\z) }\,
 \nn \\
&
\times
\frac{({\w'}^j\!-\!{\v'}^j)}{|\w'\!-\!\v'|}\, \bar{Q}\, \textrm{K}_1\left(\bar{Q}\, |\w'\!-\!\v'|\right)\, 
 \int_{-L^+/2}^{L^+/2} dz^+\, 
\Bigg\langle
\frac{1}{N_c}\, {\rm Tr} 
 \Big[\mathcal{U}_F(\w' )
\mathcal{U}_F^{\dag}(\v' )
-1\Big]
\bigg[ \mathcal{U}_F\Big(\frac{L^+}{2},z^+;\z\Big) \overleftrightarrow{ {\cal D}_{\z^j}}  \mathcal{U}_F^{\dagger}\Big(\frac{L^+}{2},z^+;\z\Big)\bigg]
 \Bigg\rangle
\, \nn
\label{Cross_Section_in_T}
\end{align}}

{\footnotesize \begin{align}
 \frac{d\sigma_{\gamma^{*}_T\rightarrow q_1\bar q_2}}{d {\rm P.S.}}\Bigg|_{\rm NEik \, corr.}^{L^+\textrm{ phase}}
=&\,
   2\pi \delta(k_1^+\!+\!k_2^+\!-\!q^+)\, N_c\, \alpha_{\textrm{em}}\,  
  e_f^2 \, 
 \left[1+\left(\frac{k_2^+\!-\!k_1^+}{q^+}\right)^2\right]\, 
2 {\rm Re}\;  (-i)\, \frac{L^+}{2}\;\int_{\z, \v', \w'}
e^{i \k_1 \cdot (\v'-\z) } \, 
 \nn \\
&
\hspace{-3.5cm}\times e^{i\k_2 \cdot (\w'-\z) }\,
\frac{({\w'}^j\!-\!{\v'}^j)}{|\w'\!-\!\v'|}\, \bar{Q}\, \textrm{K}_1\left(\bar{Q}\, |\w'\!-\!\v'|\right)\, 
\Bigg\langle
\frac{1}{N_c}\, 
{\rm Tr} 
 \Big[\mathcal{U}_F(\w' )
\mathcal{U}_F^{\dag}(\v' )
-1\Big]
\Big[\mathcal{U}_F(\z)\overleftrightarrow{\partial_{\z^j}}
\mathcal{U}_F^{\dag}(\z)
\Big]
 \Bigg\rangle
\, \nn
\end{align}}

{\footnotesize\begin{align}
& \frac{d\sigma_{\gamma^{*}_T\rightarrow q_1\bar q_2}}{d {\rm P.S.}}\Bigg|_{\rm NEik \, corr.}^{\textrm{dyn. target}}
=\,
   2\pi \delta(k_1^+\!+\!k_2^+\!-\!q^+)\, 
 N_c\, \frac{\alpha_{\textrm{em}}}{\pi} \, e_f^2\,  
\frac{k_1^+ k_2^+\,(k_2^+\!-\!k_1^+)}{(q^+)^3}\,
2 {\rm Re}\; (-i)
  \int_{\v, \v', \w, \w'}
e^{i \k_1 \cdot (\v'-\v) } \, e^{i\k_2 \cdot (\w'-\w) }\;
\nn \\
&
\hspace{-0.2cm}\times
\bigg[\tilde Q(\w',\v',\v_*,\w_*)-\tilde d(\v_*,\w_*) \bigg]
\Bigg\{ 
 \frac{1}{2}\, \left[1+\left(\frac{k_2^+\!-\!k_1^+}{q^+}\right)^2\right]\, 
\frac{({\w'}\!-\!{\v'})\!\cdot\!(\w\!-\!\v)}{|\w'\!-\!\v'|}\, \bar{Q}\, \textrm{K}_1\left(\bar{Q}\, |\w'\!-\!\v'|\right)\, 
 Q^2\, \textrm{K}_0\left(\bar{Q}\, |\w\!-\!\v|\right)
  \nn \\
&\;\;\;\;
+m^2\, Q^2\,  \textrm{K}_0\left(\bar{Q}\, |\w'\!-\!\v'|\right)\, 
\frac{|\w\!-\!\v|}{\bar{Q}}\: 
 \textrm{K}_1\left(\bar{Q}\, |\w\!-\!\v|\right)\,
+2 \, 
 \frac{({\w'}\!-\!{\v'})\!\cdot\!(\w\!-\!\v)}{|\w'\!-\!\v'||\w\!-\!\v|}\, \bar{Q}^2\, \textrm{K}_1\left(\bar{Q}\, |\w'\!-\!\v'|\right)\, 
  \textrm{K}_1\left(\bar{Q}\, |\w\!-\!\v|\right)\,
\Bigg\}
\, \nn
\end{align}}

{\footnotesize \begin{align}
 &
 \frac{d\sigma_{\gamma^{*}_T\rightarrow q_1\bar q_2}}{d {\rm P.S.}}\Bigg|_{\rm NEik \, corr.}^{\textrm{dec. on }q}
= 
 2\pi \delta(k_1^+\!+\!k_2^+\!-\!q^+)\,  N_c\, \frac{\alpha_{\textrm{em}}}{\pi} \, e_f^2\,  
 \frac{2 k_2^+}{ q^+}\,
2 {\rm Re}\;   \int_{\v, \v', \w, \w'}
e^{i \k_1 \cdot (\v'-\v) } \, e^{i\k_2 \cdot (\w'-\w) }\;
 \nn \\
&\,
\times
\Bigg\{
\left[
 \left(\frac{ (\k_2^j\!-\!\k_1^j)}{2}\,
+\frac{i}{2}\,  \partial_{\w^j} \right)\!
\left(Q^{(1)}_j(\w',\v',\v_*,\w)\!-\!d^{(1)}_j(\v_*,\w)\right)
 -i \left(Q^{(2)}(\w',\v',\v_*,\w)\!-\!d^{(2)}(\v_*,\w)\right)
  \right]
\nn \\
&\,
\times
\Bigg[
\frac{1}{2} \left(1+\left(\frac{k_2^+\!-\!k_1^+}{q^+}\right)^2\right)
 \frac{({\w'}\!-\!{\v'})\!\cdot\!(\w\!-\!\v)}{|\w'\!-\!\v'||\w\!-\!\v|}\, \bar{Q}^2\, \textrm{K}_1\left(\bar{Q}\, |\w'\!-\!\v'|\right)
 \textrm{K}_1\left(\bar{Q}\, |\w\!-\!\v|\right)\, 
\nn \\
&\,
+ m^2\, \textrm{K}_0\left(\bar{Q}\, |\w'\!-\!\v'|\right)\, \textrm{K}_0\left(\bar{Q}\, |\w\!-\!\v|\right)
\Bigg]
\nn \\
&\,
+
 \frac{(k_1^+\!-\!k_2^+)}{q^+} \,   \frac{({\w'}^{i}\!-\!{\v'}^{i})(\w^j\!-\!\v^j)}{|\w'\!-\!\v'||\w\!-\!\v|}\, \bar{Q}^2\, \textrm{K}_1\left(\bar{Q}\, |\w'\!-\!\v'|\right)\, 
 \textrm{K}_1\left(\bar{Q}\, |\w\!-\!\v|\right)\, 
 \left(Q^{(3)}_{ij}(\w',\v',\v_*,\w)\!-\!d^{(3)}_{ij}(\v_*,\w)\right)
 \Bigg\}
\, \nn
\end{align}}

{\footnotesize \begin{align}
 \frac{d\sigma_{\gamma^{*}_T\rightarrow q_1\bar q_2}}{d {\rm P.S.}}\Bigg|_{\rm NEik \, corr.}^{\textrm{dec. on }\bar{q}}
&
=  2\pi \delta(k_1^+\!+\!k_2^+\!-\!q^+)\, N_c\, \frac{\alpha_{\textrm{em}}}{\pi} \, e_f^2\, 
  \frac{2 k_1^+}{ q^+}\,   
2 {\rm Re}
\int_{\v, \v', \w, \w'}
e^{i \k_1 \cdot (\v'-\v) } \, e^{i\k_2 \cdot (\w'-\w) }\;
\nn \\
&
\hspace{-2.8cm}
\times 
\Bigg\{
\Bigg[
\frac{1}{2}\, \left[1+\left(\frac{k_2^+\!-\!k_1^+}{q^+}\right)^2\right] 
\frac{({\w'}\!-\!{\v'})\!\cdot\!(\w\!-\!\v)}{|\w'\!-\!\v'||\w\!-\!\v|}\, \bar{Q}^2\, \textrm{K}_1\left(\bar{Q}\, |\w'\!-\!\v'|\right)\, 
  \textrm{K}_1\left(\bar{Q}\, |\w\!-\!\v|\right)\, 
\nn \\
&
\hspace{-2.8cm}+ m^2\, \textrm{K}_0\left(\bar{Q}\, |\w'\!-\!\v'|\right)\, \textrm{K}_0\left(\bar{Q}\, |\w\!-\!\v|\right)
\Bigg]
\nn \\
&
\hspace{-2.8cm}
\times 
 \bigg\{
 \Big[-\frac{(\k_2^j\!-\!\k_1^j)}{2} 
 +\frac{i}{2}\,\partial_{\v^j}
 \Big]
 \Big[{Q^{(1)}_j(\v',\w',\w_*,\v)}^{\dag}-{d^{(1)}_j(\w_*,\v)}^{\dag} \Big]
 -i  \Big[{Q^{(2)}(\v',\w',\w_*,\v)}^{\dag}-{d^{(2)}(\w_*,\v)}^{\dag}\Big]
   \bigg\}
\nn \\
&
\hspace{-2.8cm}
+
 \frac{(k_1^+\!-\!k_2^+)}{q^+}  
\frac{({\w'}^{i}\!-\!{\v'}^{i})(\w^j\!-\!\v^j)}{|\w'\!-\!\v'||\w\!-\!\v|}\, \bar{Q}^2\, \textrm{K}_1\left(\bar{Q}\, |\w'\!-\!\v'|\right)\, 
 \textrm{K}_1\left(\bar{Q}\, |\w\!-\!\v|\right)\, 
 \Big[{Q^{(3)}_{ij}(\v',\w',\w_*,\v)}^{\dag}-{d^{(3)}_{ij}(\w_*,\v)}^{\dag}\Big]
\Bigg\}
\, \nn
\end{align}} 
As in the previous case, this result is presented in terms of the decorated dipoles ($\tilde d,d^{(1)},d^{(2)}$ and $d^{(3)}_{ij}$) and quadrupoles ( $\tilde Q,Q^{(1)}_j,Q^{(2)}$ and $Q^{(3)}_{ij}$) that are defined explicitly in \cite{mipaper2}.


\end{document}